\newcount\style \style=2
\ifodd\style
 \documentstyle[aas2pp4]{article}
  \else
   \documentstyle[12pt,aasms4]{article}
\fi 

\begin{document}
\setlength{\baselineskip}{12pt}

\newcommand\bb[1] {   \mbox{\boldmath{$#1$}}  }

\newcommand\del{\bb{\nabla}}
\newcommand\bcdot{\bb{\cdot}}
\newcommand\btimes{\bb{\times}}
\newcommand\vv{\bb{v}}
\newcommand\B{\bb{B}}
\newcommand\BV{Brunt-V\"ais\"al\"a\ }
\newcommand\iw{ i \omega }
\newcommand\kva{ \bb{k\cdot v_A}  }
\newcommand\kb{ \bb{k\cdot b}  }
\newcommand\kkz { \left( \frac{k}{k_Z}\right)^2\>}

    \def\dd{\partial}
    \def\tilde{\widetilde}
    \def\etal{et al.}
    \def\eg{e.g. }
    \def\etc{{\it etc.}}
    \def\ie{i.e.}
    \def\beq{ \begin{equation} }
    \def\eeq{ \end{equation} }
    \def\spose#1{\hbox to 0pt{#1\hss}} 
    \def\ltsim{\mathrel{\spose{\lower.5ex\hbox{$\mathchar"218$}}
	 \raise.4ex\hbox{$\mathchar"13C$}}}

\def\tilde{\widetilde}

\newcommand{\schwz}{ {\dd  \ln P\rho ^{-5/3} \over \dd Z}}
\newcommand{\schwR} { {\dd  \ln P\rho ^{-5/3} \over \dd R} }

\newcommand{\balbz}{ {\dd  \ln T \over \dd Z}}
\newcommand{\balbR} { {\dd  \ln T \over \dd R} }

\long\def\Ignore#1{\relax}

\title{Convective and Rotational Stability of a Dilute Plasma}

\author{Steven A.~Balbus}
\affil{Virginia Institute of Theoretical Astronomy, Department of Astronomy,}
\affil{University of Virginia, Charlottesville, VA 22903-0818}
\affil{sb@virginia.edu}

\vskip 2 truein
\vskip 2 truein

\begin{abstract}

The stability of a dilute plasma to local convective and rotational
disturbances is examined.  A subthermal magnetic field and finite
thermal conductivity along the field lines are included in the
analysis.  Stability criteria similar in form to the classical H\o
iland inequalities are found, but with angular velocity gradients
replacing angular momentum gradients, and temperature gradients
replacing entropy gradients.  These criteria are indifferent to the
properties of the magnetic field and to the magnitude of the thermal
conductivity.  Angular velocity gradients and temperature gradients are
both free energy sources; it is not surprising that they are directly
relevant to the stability of the gas.  Magnetic fields and thermal
conductivity provide the means by which these sources can be tapped.
Previous studies have generally been based upon the classical H\o iland
criteria, which are inappropriate for magnetized, dilute astrophysical
plasmas.  In sharp contrast to recent claims in the literature, the new
stability criteria demonstrate that marginal flow stability is not a
fundamental property of accreting plasmas thought to be associated with
low luminosity X-ray sources.


\end{abstract}

\section{Introduction}


Accretion onto compact objects is generally possible only if specific
angular momentum is somehow extracted from fluid elements.  It is now
known that magnetic fields in a differentially rotating fluid cause a
breakdown of laminar flow into turbulence (Balbus \& Hawley 1998), and
that this turbulence leads to a much enhanced angular momentum
transport.  Despite the fact that turbulence is to some extent amenable
to detailed numerical simulation, the theory of turbulent transport
lacks a good phenomenological description.  It may well be that one
does not exist.

To make progress, some form of idealization is usually necessary, and a
common approach has been the following: the magnetohydrodynamic (MHD)
turbulent transport is modeled as an enhanced Navier-Stokes viscosity
in an {\em unmagnetized} fluid.  The idea is that the flow of this
ansatz hydrodynamical fluid may then be subject to further
instabilities.  In particular, the enhanced viscosity is also an energy
source, and heating in the inner regions of some types of accretion
flow leads, it is argued, to a convectively-unstable temperature
gradient.  It has been further argued that, much like stellar
convective zones, the flow would quickly evolve to a state of marginal
stability.  Proponents of these convection-dominated accretion flows
(`CDAFs') suggest that this process explains the very low X-ray
luminosities associated with many black-hole candidates (Narayan,
Igumenschev, \& Abramowicz 2000), because the {\em inward} angular
momentum transport associated with thermal convection would precisely
offset the usual outward MHD transport.  In the stellar case, marginal
stability amounts to adopting an adiabatic temperature profile; in an
accretion flow matters are more complex, with constant entropy and
angular momentum surfaces satisfying marginal stability by the H\o
iland criterion (Quataert \& Gruzinov 2000).

We are therefore motivated to consider the equilibrium of a rotating,
magnetized, hot dilute plasma.  The results of this investigation may
be surprising to the reader, because they are at odds with what has
become a standard approach.  Our findings are relevant to understanding
current problems in X-ray accretion sources, but they are also of
general fluid dynamical interest.  They are easily stated: (1) In a
rotating, stratified, dilute plasma, the presence of any magnetic field
and any Coulomb-based thermal conductivity renders the classical H\o
iland criteria insufficient for flow stability (to axisymmetric
disturbances).   Maximum growth rates are rapid (dynamical time scale),
and insensitive to the field strength and the value of the
conductivity.  (2) Where entropy ($S$) and angular momentum ($l$)
gradients appear in the classical H\o iland formulae, they must be
replaced with temperature ($T$) and angular velocity ($\Omega$)
gradients.  {\it Spherical adiabatic accretion is unstable.}  (3)
Convective instability is inseparable from rotational instability, in
the sense that the breakdown of laminar flow into turbulent eddies and
the formation of convective eddies are inseparable processes, governed
by identical intertwined criteria.  Marginal stability is no more
likely to be attained in advection-dominated type flows than it is in
disks.  Indeed, it is the departure from marginal stability that is
critical for maintaining vigorous turbulent transport, transport which
is at once affected by the dynamical and the thermal properties of the
flow.

The following section presents a detailed development of these results,
and the final section of this paper discusses some the consequences
for understanding the stability of hot plasma winds and accretion
flows.

\section{Local Stability of a Dilute Plasma}

\subsection{Physical Description}

The appearance of $\Omega$ and $T$ gradients as stability discriminants
has fundamental significance.  They strongly destabilize classically
stable flows.  Before entering a technical discussion, it is desirable
to have a physical understanding of how this arises in some illustrative
cases.

In figure (1), we contrast the stability behavior of rotating fluid
elements in an unmagnetized (1a) and magnetized (1b) medium.  In the
unmagnetized case, the fluid element retains its angular momentum $l_1$
on its epicyclic excursion.  The angular momentum of an undisturbed
orbit at the new location is $l_2$.  The Rayleigh criterion for
stability is simply $l_1<l_2$, so that the displaced element would drop
back to a lower orbit.  This means that the angular momentum in the
disk increases outward.  The magnetized medium is distinguished by
magnetically tethered fluid elements, which enforces isorotation at
marginal stability, i.e., $\Omega_1 = \Omega_2$.  Now the criterion for
stability (angular momentum less than surroundings) is
$\Omega_1<\Omega_2$.  Hence, a magnetized disk requires that the
angular velocity, not the angular momentum, increase outward for
stability.  This is the magnetorotational stability criterion.

Figure (2) shows the configuration appropriate to a thermally
stratified medium.  In the unmagnetized case (2a), the fluid element
retains its entropy $S_1$ as it is displaced.  The entropy of the
undisturbed layer at the element's new location is $S_2$.  The
Schwarzschild criterion for stability is $S_1<S_2$, which is just the
condition for the blob to be denser than its (pressure-equilibrium)
surroundings.  This means the entropy increases upwards.

In the magnetized case (2b), the displaced fluid element draws magnetic
field lines up with it, and the temperature gradient now has a
component along the displaced field line.  Heat flows along this
direction from the hotter to the cooler element.  Under conditions of
marginal stability, thermal conduction maintains a constant temperature
between the fluid elements.  The criterion for buoyant stability is now
$T_1<T_2$.  A magnetized medium requires that the temperature, not the
angular momentum, increase upward for stability.  We shall now show
that the classical H\o iland criteria are simply but significantly
modified to accommodate these new stability criteria.

%

\subsection {Analysis}

In the presence of a magnetic field, heat is restricted to flowing 
along lines of magnetic force, provided that the ion Larmor radius
is much less than the collisional mean free path
(e.g. Braginskii 1965).  This is equivalent to
\beq
\omega_{cI}\tau \gg 1
\eeq
where $\omega_{cI}$ is the ion cyclotron frequency, and $\tau$ is the
mean free collision time.  This inequality is generally satisfied
by dilute astrophysical plasmas.  Under these circumstances, the
electron heat conduction parallel to the magnetic field is given by
\beq
\bb{Q} = -\chi \bb{b}\, (\bb{b\cdot\nabla}) T_e
\eeq
where $\chi$ is the electron conductivity (Spitzer 1962),
\beq
\chi \simeq 6\times 10^{-7} \, T_e^{5/2} \ {\rm ergs\> cm^{-1}\>
K^{-1}}\, ,
\eeq
and $T_e$ is the electron temperature.   
$\bb{b}$ is a unit vector in the direction of the magnetic field.
Unless otherwise explicitly stated, we shall assume that the ions
and electrons have the same temperature $T$. 
This amounts to requiring
that the mean free path $\lambda$ and the flow scale height $H$
satisfy $\lambda/H \ll 1$ (Cowie and McKee 1977), or
\beq
10^4 {T^2\over n H} \ll 1.
\eeq

The usual fluid equations for a monotomic plasma are
\beq
{\dd\rho\over \dd t} + \del\bcdot (\rho \bb {v}) =  0,
\eeq
\beq\label{mom}
\rho {\dd\vv \over \dd t} + (\rho \vv\bcdot\del)\vv = -\del\left(
P + {B^2\over 8 \pi} \right)-\rho \del \Phi +
\left( {\B\over 4\pi}\bcdot \del\right)\B,
\eeq
\beq
{\dd\bb{B}\over \dd t} = \bb {\nabla\times (v\times B)},
\eeq
\beq
{3\over 2} {P} {d\ln P\rho^{-5/3}\over dt} = - \del\bcdot\bb{Q}
=  \del\bcdot\left[ \bb{b}\left(\chi \bb{b}\bcdot\del T\right)\right].
\eeq
These are respectively mass conservation, the equation of motion, the
induction equation, and the entropy equation.   Our notation is
standard: $\bb{v}$ is the fluid velocity, $\rho$ the mass density, $P$
the gas pressure, $\bb{B}$  the magnetic field, and $\Phi$ the
gravitational potential.  The ion-dominated viscosity is ignored, as it
is smaller than the conductivity by a factor of order the square root
of the electron-to-ion mass ratio.  (This assumes equal ion and
electron temperatures.)  The plasma is taken to be a perfect
electrical conductor.

We assume that in the equilibrium solution, the field lines are
isotherms.  Then the linearly perturbed heat flux is
\beq
\bb{\delta Q} = -\chi \bb{b}\,  (\bb{\delta b\cdot \nabla}) T -
\chi\bb{b} (\bb{b\cdot \nabla}) \delta T.
\eeq
Perturbed quantities have the WKB space-time dependence
$\exp i(\bb{k\cdot r} - \omega t)$, where $\bb{k}$ is the
axisymmetric wavevector $(k_R, 0, k_Z)$.
We shall work in the Boussinesq limit, and the linearly perturbed
equations are therefore
\beq
\del\bcdot\bb{\delta v} = 0,
\eeq
\beq
{\dd\delta\bb{v}\over \dd t} + \delta \bb{v}\bcdot\del\bb{v} =
{\delta\rho\over \rho^2} \bb{\nabla} P
-{1\over\rho} \del\left( \delta P + { \bb{\delta B\cdot B}\over
4\pi}\right) + { (\bb{B}\bcdot\nabla) \over4\pi\rho} \bb{\delta
B},
\eeq
\beq
{\dd \bb{\delta B}\over \dd t} = \bb{\nabla\times (\delta v\times
B)} + \bb{\nabla\times (v\times \delta B)},  
\eeq
\beq
{5\over 3} {\dd\ \over\dd t} {\delta\rho\over\rho} - \bb{\delta
v\cdot\nabla}\ln P\rho^{-5/3} = {2\over
3P}\bb{\nabla\cdot\delta Q} .
\eeq
In explicit component form, the leading order WKB equations are
\beq\label{divv}
k_R \delta v_R + k_Z \delta v_Z =0,
\eeq
\begin{eqnarray}
-\iw \delta v_R && +{i k_R \over \rho} \delta P - 2 \Omega \delta
v_\phi
- {\delta\rho\over\rho^2} { \dd P \over \dd R} + {i k_R \over 4
\pi
\rho}\nonumber\\
&&\times \left( B_\phi \,\delta B_\phi + B_Z \,\delta B_Z
\right) - {ik_Z\over
4 \pi\rho}B_Z\, \delta B_R = 0,
\end{eqnarray}
\beq
-\iw \delta v_\phi + \delta v_R\, {1\over R}{\dd (R^2\Omega) \over \dd R}+
\delta v_Z\, R{\dd \Omega\over \dd Z}
-i \bb{k \cdot B} {\delta B_\phi \over 4 \pi \rho} = 0,
\eeq
\begin{eqnarray}
-\iw \delta v_Z &&+ {i k_Z \, \delta P \over \rho} -
{\delta\rho\over
\rho^2}{ \dd P \over \dd Z} + {i k_Z \over 4 \pi \rho
}\nonumber\\
&& \times \left( B_\phi \, \delta B_\phi
+ B_R\, \delta B_R \right) - {ik_R B_R\over 4 \pi \rho}\, \delta
B_Z =0,
\end{eqnarray}
\beq\label{indR}
-\iw \delta B_R - i \bb{k \cdot B} \delta v_R = 0,
\eeq
\beq
-\iw \delta B_\phi - \delta B_R\, {\dd\Omega\over \dd \ln R} -
\delta
B_Z\, R {\dd\Omega\over \dd Z} - i \bb{k\cdot B} \, \delta
v_\phi = 0,
\eeq
\beq\label{indZ}
-\iw \delta B_Z - i \bb{k\cdot B} \delta v_Z = 0,
\eeq
\beq\label{lincon}
\iw {5\over3} {\delta\rho\over \rho} +
\bb{(\delta v\cdot\nabla)}\ln P\rho^{-5/3} =
{2\chi\over 3P}\left(
i  \bb{k\cdot b}\,  (\bb{\delta b\cdot \nabla}) T -
\bb{(k\cdot b)}^2 \delta T \right).
\eeq
Note that the change in the unit vector $\bb{b}$ is
\beq\label{delb}
\bb{\delta b} = \bb{b\times}\left(  { \bb{\delta B}\over {B}}
\bb{\times b} \right).
\eeq

These equations differ from the {\em adiabatic} magnetized H\o iland
criteria studied by Balbus (1995, hereafter B95)
only in the final energy equation
(\ref{lincon}), which contains a rather complicated-looking conduction
term.  Despite the apparent complexity, the fact that only the energy
equation has changed allows us to simplify considerably our calculation
of the dispersion relation, by making use of the B95 result.

Begin by setting $\delta T = - T(\delta\rho/\rho)$ in the conduction
term, which is an implementation of the Boussinesq approximation
(relative changes in the pressure are much smaller than relative
changes in the temperature or density).  Then, substituting for
$\bb{\delta b}$ from equation (\ref{delb}), and remembering that the
field lines are isothermal, brings us to
\beq
\left( \iw - {2\over5}{\chi T\over P} \bb{ (k\cdot b)}^2 \right)
{\delta\rho\over \rho} = - {3\over 5} \bb{\delta v \cdot\nabla}
\ln P\rho^{-5/3} + {2i\chi T\over 5P} \bb{(k\cdot b)} {
\bb{\delta B} \over B} \bb{\cdot\nabla}\ln T.
\eeq
Using equations (\ref{indR}), (\ref{indZ}) 
and (\ref{divv}), then simplifying, leads to
\beq\label{crux}
{\delta \rho \over \rho} = 
{\delta v_R\over \iw} \  \left[ \frac {
\displaystyle\frac{3}{5}
{\cal D} \ln P\rho^{-5/3} + \frac {2\chi T i} {5 P
\omega} \bb{(k\cdot b)}^2 {\cal D}\ln T} 
{\displaystyle
1 +  \frac {2\chi T i}{ 5 P\omega}\bb{(k\cdot b)}^2
} \right] \equiv {\delta v_R\over \iw}{3\Theta\over5}
{\cal D} \ln P\rho^{-5/3},
\eeq
where 
\beq
{\cal D} = \left( 
{k_R\over k_Z}{\dd\ \over \dd Z} - {\dd\ \over\dd R}
\right),
\eeq
and $\Theta$ is defined {\em in situ.}

Equation (\ref{crux}) has two important limiting forms.  When 
$\chi\rightarrow 0$ and $\Theta\rightarrow 1$,
we recover the adiabatic expression
\beq
\label{adiab} {\delta \rho \over \rho} = {\delta v_R\over \iw} \
\frac{3}{5} {\cal D} \ln P\rho^{-5/3} =
 - \xi_R\ \frac{3}{5} {\cal D} \ln P\rho^{-5/3}, \eeq
where we have introduced the radial Lagrangian fluid
displacement, $\delta v_R = d\xi_R /dt$.  This returns the
calculation to B95.  
But note the result obtained if
we first take the limit $\omega \rightarrow 0$, followed by
$\chi\rightarrow 0$.  Then $\Theta$ is proportional to the ratio of the
temperature-to-entropy ${\cal D}$ gradients, and
\beq\label{nadiab}
{\delta \rho \over \rho} = -\xi_R\  {\cal D} \ln T.
\eeq
This implies that the $\omega\rightarrow 0$ limit of the dispersion
relation is simply obtained by taking the B95 result and replacing
$(3/5){\cal D}\ln P\rho^{-5/3}$ with ${\cal D}\ln T$.  Since the
$\omega\rightarrow 0$ limit is relevant for flow stability, comparison
of equations (\ref{adiab}) and (\ref{nadiab}) points to something
remarkable:  the stability of the flow changes discontinuously when any
finite thermal conductivity is present.  Instead of entropy gradients
serving as the discriminant for buoyant stability, temperature
gradients are key.  If the temperature --- {\em not} the entropy ---
decreases upwards, the plasma will become buoyantly unstable.  This is
a very big difference indeed, implying that simple adiabatic
stratification (or flow) is unstable (Balbus 2000).  It is wholly
analogous to the replacement of angular momentum gradients with angular
velocity gradients in a magnetized rotator, and the accompanying
destabilization of Keplerian flow.

\subsection{Dispersion Relation}

The above reasoning suggests a simple way to obtain the desired
dispersion relation: use $\Theta$ as a ``tag,'' and substitute
$3\Theta/5$ for $3/5$ where it appears as a prefactor in equation
(2.4) in B95.  This prescription gives directly
\
$$
{\tilde\omega}^4 \, {k^2\over k_Z^2} + {\tilde\omega}^2
\left[ {3\Theta\over5\rho}\, \left({\cal D} P\right)\, {\cal D} \ln
P\rho^{-5/3}
+
{1\over R^3}\, {\cal D} (R^4\Omega^2) \right] - 4 \Omega^2 (\kva)^2 =
0,
$$
where 
\beq
\bb{v_A} = { \bb{B}/\sqrt{4\pi\rho}}, 
\qquad \tilde\omega^2 = \omega^2 -(\kva)^2, \qquad 
k^2 = k_R^2 + k_Z^2.
\eeq
We may now expand this equation using (\ref{crux}), introducing
$\sigma = - i\omega$ to keep the dispersion coefficients real.  
One finds:
\begin{eqnarray}
{k^2\over k_Z^2} \tilde\sigma^4 \sigma_{cond}
- \tilde\sigma^2 \bigg[ {3\over 5\rho} {\cal D} P 
\bigg(\sigma {\cal D}\ln P\rho^{-5/3}+ 
{2\chi T\over 3P}(\kb)^2{\cal D}\ln T\bigg) + \nonumber \\
+ {1\over R^3} {\cal D}(R^4\Omega^2)
\sigma_{cond}\bigg]
- 4 \Omega^2(\kva)^2\sigma_{cond}  = 0,
\end{eqnarray}
where now
\beq
\tilde\sigma^2 =\sigma^2 + (\kva)^2, \quad \sigma_{cond} = \sigma
+ {2\chi T\over 5P}(\kb)^2
\eeq
This is the general form of the dispersion relation.  

The marginal stability of purely evanescent modes may be studied very
simply by passing 
through the point $\sigma \rightarrow 0$.  The condition for stability
is
\beq\label{stbcond}
(\kva)^2{k^2\over k_Z^2}  -{1\over\rho}({\cal D}P) ({\cal D}\ln T)
-{1\over R^3} {\cal D}(R^4\Omega^2) - 4\Omega^2 > 0.
\eeq
Note that this condition is independent of the conductivity $\chi$,
for any finite value of this parameter.
By way of contrast, when $\chi = 0$, the  $\sigma\rightarrow 0$ result is
\beq
(\kva)^2{k^2\over k_Z^2}  -{3\over5\rho}({\cal D}P) ({\cal D}\ln
P\rho^{-5/3}) -{1\over R^3} {\cal D}(R^4\Omega^2) - 4\Omega^2 > 0,
\eeq
which is the same as the previous condition,
except for the replacement of a logarithmic temperature gradient
by an entropy gradient.  

The condition (\ref{stbcond}) is actually much more general than the above
simple derivation would suggest.  It is in fact a necessary and sufficient
condition to preclude both instability and overstability.  This is 
established in detail in the Appendix by use of the Routh-Hurwitz criterion.  

At this point of the calculation, the route becomes identical to B95,
and the stability criteria are obtained by direct substitution of
${\cal D}\ln T $ for $(3/5){\cal D}\ln P\rho^{-5/3}$ in equations (2.9)
and (2.11) of that paper.  For ease of reference and cross comparison,
we give three forms of the H\o iland stability criteria, and the
conditions under which they are valid:

\noindent The textbook case (Tassoul 1978) is adiabatic and unmagnetized.

\noindent CLASSICAL H\O ILAND CRITERIA.
\beq\label{hoilad1}
-{3\over 5\rho}(\del P)\bcdot\del\ln P\rho^{-5/3}
+ {1\over R^3} {\dd R^4\Omega^2\over \dd R} \ge 0,
\eeq
\beq\label{hoilad2}
\left( - {\dd P\over \dd Z} \right) \, \left(
{\dd R^4 \Omega^2\over\dd R} \schwz - {\dd R^4\Omega^2\over\dd Z}\schwR
\right) \ge 0.
\eeq
The B95 result allows for the presence of a weak magnetic field, but ignores
thermal conduction.

\noindent ADIABATIC, MAGNETIZED CRITERIA.
\beq\label{hoilb1}
-{3\over 5\rho}(\del P)\bcdot\del\ln P\rho^{-5/3}
+ {\dd \Omega^2\over \dd \ln R} \ge 0,
\eeq
\beq\label{hoilb2}
\left( - {\dd P\over \dd Z} \right) \, \left(
{\dd \Omega^2\over\dd R} \schwz - {\dd \Omega^2\over\dd Z}\schwR
\right) \ge 0.
\eeq
The result of this paper includes both the dynamics of a weak magnetic field,
and the effects of magnetically inhibited Coulomb conductivity.

\noindent NONADIABATIC, MAGNETIZED CRITERIA (This paper).
\beq\label{hoil1}
-{1\over \rho}(\del P)\bcdot\del\ln T
+ {\dd \Omega^2\over \dd \ln R} \ge 0,
\eeq
\beq\label{hoil2}
\left( - {\dd P\over \dd Z} \right) \, \left(
{\dd \Omega^2\over\dd R} \balbz - {\dd \Omega^2\over\dd Z}\balbR
\right) \ge 0.
\eeq

The combined effect of a magnetic field and Coulomb conductivity is to
ensure that free energy sources (angular velocity and temperature
gradients) control flow stability.  This result goes some way toward
understanding why minimal energy and maximal entropy states of bound
systems are associated with uniform rotation and isothermality, yet the
classical dynamical stability criteria involve gradients of angular
momentum and entropy.  Departures from uniform rotation and
isothermality are indeed a source of dynamical instability.  It is just
that magnetic tension and magnetically confined conduction are needed
to provide the right coupling to tap into these sources.

\subsection{Some additional points}

\subsubsection{Radiative conduction}

Because of the apparent generality of the criteria (\ref{hoil1}) and
(\ref{hoil2}), it is important to emphasize the point (Balbus 2000)
that the destabilizing role of thermal conduction is modified
substantially when diffusivity is dominated by radiative processes, as
in stellar interiors.  This form of the heat conduction is indifferent
to the magnetic field.  Denoting the radiative conductivity as
$\chi_{rad}$, we have (Schwarzschild 1958):
\beq
\chi_{rad} = {16T^3\sigma\over 3 \kappa\rho},
\eeq
where $\sigma$ is the Stefan-Boltzmann constant and $\kappa$ is the 
radiative opacity.   
When both Coulomb and radiative conductivity are present,
the right hand side of equation (\ref{lincon}) is modified to
\beq
{2\over 3P}\left[
i  \chi \bb{k\cdot b}\,  (\bb{\delta b\cdot \nabla}) T -
\left(\chi \bb{(k\cdot b)}^2 +  \chi_{rad} k^2\right)  \delta T \right].
\eeq
(We assume, as before, that unperturbed field lines are isothermal.
Our ultimate conclusion is not strongly affected by this assumption.)
The dispersion relation becomes
\begin{eqnarray}
{k^2\over k_Z^2}
\tilde\sigma^4\sigma_{rad}
- \tilde\sigma^2 \bigg[ {3\over 5\rho} {\cal D} P
\bigg(\sigma {\cal D}\ln P\rho^{-5/3}+
{2\chi T\over 3P}(\kb)^2{\cal D}\ln T\bigg) + \nonumber \\
+ {1\over R^3} {\cal D}(R^4\Omega^2)\sigma_{rad} \bigg]
- 4 \Omega^2(\kva)^2\sigma_{rad} = 0,
\end{eqnarray}
where
\beq
\sigma_{rad} = \sigma + {2T\over 5P} (\chi(\kb)^2+\chi_{rad} k^2).
\eeq
The stability condition (\ref{stbcond}) becomes
\beq
(\kva)^2{k^2\over k_Z^2}  -\frac { \displaystyle
(1/\rho)({\cal D}P) ({\cal D}\ln T)} {\displaystyle  1 + (\chi_{rad}/\chi)
(k/\kb)^2 }
-{1\over R^3} {\cal D}(R^4\Omega^2) - 4\Omega^2 > 0.
\eeq
By reducing the effective size of the potentially destabilizing
temperature gradient, radiative conductivity is strongly stabilizing.
Even when formally present, instability occurs primarily at long
wavelengths (along the field lines) when radiative conduction is
dominant, and the local WKB approximation we have been using breaks
down.  When unstable wavelengths are calculated to be in excess of the
size of the system, the question of stability must be determined by a
global analysis.  The dynamical stability of stellar radiative
interiors, therefore, is not threatened by this analysis (Balbus
2000).

\subsubsection{Instability in a dynamical background}

The development of a local instability in an evolving, dynamically
active background is more complex than the analogous problem
for a static equilibrium (e.g. Balbus \& Soker 1989).  Both accretion
flows and winds are a natural venue for the instabilities of interest,
but we shall defer a detailed study of the dynamical linear stability
theory to a separate investigation (Balbus 2001, in preparation).
It is nevertheless both possible and useful
to make some simple statements of a general nature.
We restrict our comments to spherical flow.  

At $t=0$, label each fluid element by a position vector $\bb{r'}$,
which is a Lagrangian coordinate system, comoving with the unperturbed flow.
The position vector $\bb{r}$ is the instantaneous Eulerian coordinate
of a fluid element as it flows.  Radial stretching of the flow is characterized
by the scale factor
$$
a(t) \equiv {\dd r\over \dd r'},
$$
which is a sort of Hubble parameter.  The fundamental perturbation
variable in a dynamical flow is $\xi_r/a$, where $\xi_r$ is the radial
displacement of a perturbed fluid element relative to the undisturbed flow.
Note that we normalize $\xi_r$ relative to $a$, since the latter tracks the 
relative separation of two points in the underlying flow.  Displacements relative
to $a$ track true, physical changes.  

If the growth rate of an embedded instability is rapid compared with the background
flow evolution, the resulting form for $\xi_r/a$ is rather intuitive:
\beq\label{growth}
\frac{\xi_r}{a} \propto \exp\left( \int^t \gamma(t')\, dt'\right),
\eeq
where $\gamma$ is the value of the instantaneous growth rate obtained
by ignoring background flow at the fluid element's position at time
$t'$.  In a spherical system, it is not difficult to show that the most
rapidly growing thermal mode corresponds to
\beq
\gamma^2 = - {1\over \rho} {\dd P\over \dd r}{\dd\ln T\over \dd r},
\eeq
which is likely to be of order the sound crossing time.  In other
words, when the background flow is subsonic, one expects only small
quantitative changes in the development of a perturbation from static
theory.

The breakdown starts to occur when the flow becomes sonic.  Then, there
is only one time scale in the problem, and the integral in
(\ref{growth}) becomes logarithmic, with power law behavior for the
perturbations.  Once the flow becomes highly supersonic, there is a
complete breakdown of the WKB form (\ref{growth}).  Generally, such
disturbances are ``inflated away,'' and in any case cannot make contact
with upstream fluid.  Applications of the dispersion relation or the
stability criteria to dynamically active flows should be restricted to
their subsonic zones.

\section{Discussion}

The understanding that temperature and angular velocity gradients
regulate flow stability in magnetized dilute plasmas has important
consequences.  When rotation is unimportant, isothermal, not adiabatic,
conditions should prevail.  This should be true whether or not 
field lines are free to open up.  
Marginal stability is a likely outcome,
since the temperature profile is free to adjust itself while maintaining
hydrostatic equilibrium.  One interesting application is to X-ray gas
in early-type galaxies.  Fabian et al. (1986) show that the classical
Schwarzschild stability criterion implies a minimum mass contained
within a confining outer radius $r_c$ of
\beq
M \ge {5c_0^2 r_0\over2G}  {1 - (P_c/P_0)^{2/5}\over 1 - r_0/r_c},
\eeq
where $c_0$ is the isothermal sound speed at observed radius $r_0$,
at the pressures $P_c$ and $P_0$ are those at $r_c$ and $r_0$.  
This constraint may be further tightened with the isothermal criterion to
\beq
M \ge {c_0^2 r_0\over G}  {\ln (P_0/P_c)\over 1 - r_0/r_c},
\eeq
which also has the added advantage of being rather insensitive to the pressures. 
(Note that classical cooling flows are not affected by this
instability, as the temperature increases outward when radiative losses
are important.)

When application is made to a rotating systems, the notion of marginal
stability is often inappropriate, for the simple reason that the system
may not have the option of changing its rotation law.  In Keplerian
disks, the instability  reduces to the magnetorotational instability,
and that is hardly a marginally stable flow.  More contemporary
examples are ADAFs (Narayan, Mahadevan, \& Quataert 1998 for a review),
and CDAFs (Narayan et al.~2000), which were mentioned briefly in the
Introduction.  The latter are particularly striking, as proponents
argue that convection itself is capable of suppressing accretion, i.e.
that inward convective angular momentum very nearly cancels the outward
transport of MHD turbulence.  This is a rather bold extrapolation from
earlier numerical simulations (Stone \& Balbus 1996, Cabot 1996), which
suggest an effective $\alpha$ value some three orders of magnitude
smaller than that obtained from the magnetorotational instability.
Convective turbulence tends to create velocity-temperature correlations
that (not surprisingly) do a fine job of transporting heat, and a very
poor job of transporting angular momentum.  Recent MHD simulations of
resistively-heated, non-radiative, quasi-spherical accretion flows (Stone
\& Pringle 2001) show no evidence of convective stalling.

Part of the difficulty of ascribing a special role to convective angular
momentum transport may be grasped from the form of the stability
criteria appropriate to these flows, equations (\ref{hoil1}) and
(\ref{hoil2}).  CDAF proponents use, incorrectly, the classical H\o
iland criteria (\ref{hoilad1}) and (\ref{hoilad2}).  Among other
problems, this has the effect of masking the dominant magnetorotational
instability, which ultimately governs the nature of the angular
momentum transport.  The formal epicyclic frequency (the term
proportional to the angular momentum gradient in [\ref{hoilad1}]) is
real-valued in the flows of interest, and gives no hint of rotational
instability.  Only the adverse entropy gradients, which are clearly
associated with convection, would seem to be involved with
instability.  But this is very misleading.

The appropriate form of the instability criteria (\ref{hoil1}) and
(\ref{hoil2}) tell a much different story.  The point that the
thermally driven component is governed by temperature, not entropy,
gradients is a relatively minor one in this context.  (For this reason,
we need not concern ourselves here with the possible complexities of a
two-temperature plasma.)  Much more important is the conceptual point
that convective and rotational instability must be treated on the same
footing.  There is no separate ``viscous'' angular momentum transport
and ``convective'' angular momentum transport, any more than there is
turbulent transport and magnetic transport.  There is a rotationally
driven MHD instability that gives rise to a turbulent stress tensor,
and there is nothing marginally unstable about the flow.  As mentioned
above, Stone \& Pringle (2001) have performed two-dimensional
calculations that bear this out; preliminary three-dimensional
simulations are equally supportive (Hawley, Balbus, \& Stone 2001,
in preparation).

It is, nevertheless, of great interest to understand the MHD {\em
thermal} properties of nonrotating (or uniformly rotating) flows.  What
are the numerical prospects for studying the thermoclinic instability
in nonrotating systems?  It is a daunting task, because thermal
conduction along magnetic field lines must be isolated, uncontaminated
by numerical cross-field diffusion.  When the field becomes highly
tangled, computations become prohibitive.

Even at this early stage of inquiry, something can be done.  The linear
instability in a simple vertically stratified box with a weak
horizontal field can be demonstrated under Schwarzschild stable
conditions (Stone 2001, private communication).  It is even possible to
see the early stages of nonlinear instability, as shown in figure [3].
If the effect of the instability is generally to comb field lines out,
then it may well be possible to make further progress on the numerical
front.  Either by field line combing or by convection, the tendency
toward isothermality of a bottom-hot dilute stratified plasma seems a
likely outcome.

The results of this paper should make very clear the important
conceptual point that even the tiniest of magnetic fields can have
dramatic consequences for the macroscopic stability of astrophysical
plasmas.  Classical hydrodynamical results can be qualitatively
incorrect, and great care must be taken before uncritically taking them
over into magnetized systems.  Subtlety need not imply great
complexity:  the relative simplicity of our fundamental results,
equations (\ref{hoil1}) and (\ref{hoil2}), is encouragement that the
most important MHD stability properties can also be conceptually
simple.  And since it seems certain that the final word has not
been said on the topic, more surprises are likely.

It is a pleasure to thank J.~Hawley and C.~Terquem for detailed
comments on a preliminary draft of this paper which led to significant
improvements in the presentation, and M.~Fall for pointing out to me
that the new stability criteria sharpen dark matter constraints in X-ray
clusters.  I am also most grateful to J.~Stone for providing me with
fig.~[3], and for discussions related to the numerics of magnetized
thermal conduction.  This research was supported by NASA grants NAG
5--7500, NAG5-106555, NAG5-9266, and NSF grant AST00-70979.

\section*{Appendix: Necessary and Sufficient Criterion for Stability}

To prove that the condition (\ref{stbcond}) is a necessary and sufficient
condition for any type of instability, we first expand the polynomial in
equation (29) in the form
\beq\label{poly}
a_0\sigma^5 + a_1\sigma^4 + a_2\sigma^3 + a_3\sigma^2 + a_4\sigma +a_5 =0.
\eeq
where
\beq
a_0 = k^2/k_Z^2,
\eeq
\beq
a_1 = \frac{2\chi T}{5P} \frac{k^2}{k_Z^2} (\bb{k\cdot b})^2,
\eeq
\beq
a_2 = 2(\bb{k\cdot v_A})^2 \frac{k^2}{k_Z^2} - \frac{3}{5\rho}{\cal D}P\>  {\cal D}\ln P\rho^{-5/3}
- \frac{1}{R^3}{\cal D} (R^4\Omega^2),
\eeq
\beq
a_3 =  \frac{2\chi T}{5P} (\bb{k\cdot b})^2\left[
2(\bb{k\cdot v_A})^2 \frac{k^2}{k_Z^2} - \frac{1}{\rho}{\cal D}P\>  {\cal D}\ln T
- \frac{1}{R^3}{\cal D} (R^4\Omega^2)\right],
\eeq
\beq
a_4= (\bb{k\cdot v_A})^2\left[
(\bb{k\cdot v_A})^2\frac{k^2}{k_Z^2} - \frac{3}{5\rho}{\cal D}P\>  {\cal D}\ln P\rho^{-5/3}
- \frac{1}{R^3}{\cal D} (R^4\Omega^2) -4\Omega^2\right],
\eeq
\beq
a_5=
\frac{2\chi T}{5P} (\bb{k\cdot b})^2   (\bb{k\cdot v_A})^2
\left[
(\bb{k\cdot v_A})^2 \frac{k^2}{k_Z^2} - \frac{1}{\rho}{\cal D}P\>  {\cal D}\ln T
- \frac{1}{R^3}{\cal D} (R^4\Omega^2) - 4\Omega^2
\right].
\eeq
To make things more compact and manageable, notice that
\beq
\epsilon \equiv - \frac{3}{5\rho}{\cal D}P\>  {\cal D}\ln P\rho^{-5/3}
+ \frac{1}{\rho}{\cal D}P\>  {\cal D}\ln T
= \frac{2}{5\rho P}({\cal D} P)^2 > 0,
\eeq
and if we furthermore define
\beq
\delta \equiv (k^2/k_Z^2) (\bb{k\cdot v_A})^2 +4\Omega^2 > 0,
\eeq
then we may write
\beq\label{aaa}
a_3 = \frac {2\chi T}{5P} (\bb{k\cdot b})^2 (a_2 - \epsilon), \quad
a_4 = (\bb{k\cdot v_A})^2 (a_2 - \delta), \quad a_5 =
\frac {2\chi T}{5P} (\bb{k\cdot b})^2 (\bb{k\cdot v_A})^2 (a_2 - \delta-\epsilon).
\eeq

For stability, we require that the real part of $\sigma$ must be less
than zero, i.e., that the roots of equation (\ref{poly}) must all lie in the
left complex $\sigma$ plane.  Polynomials with this property are known
as {\em Hurwitz} polynomials (e.g. Levinson \& Redheffer 1970).  If
the left side of equation (\ref{poly})
is a Hurwitz polynomial, both pure exponential instability as well
as overstability are precluded.  This will be so if and only if
\beq
\left| \begin{array} {ccccc}
 a_1 &  a_0 & 0   &   0 & 0\\
 a_3 &  a_2 & a_1 & a_0 & 0\\
 a_5 &  a_4 & a_3 & a_2 & a_1\\
 0   &    0 & a_5 & a_4 & a_3\\
 0   &    0 &  0  &  0  & a_5
\end{array}   \right|  > 0,
\eeq
and all five determinants obtained by expanding along the
diagonal, i.e.,
\beq
a_1, \qquad  \left| \begin{array} {cc}
a_1 & a_0 \\
a_3 & a_2
\end{array} \right|, \qquad {\rm etc.}
\eeq
are each individually $> 0$ (Levinson \& Redheffer 1970).  
We shall denote the determinant of each $n\times n$ matrix
thus obtained as $\det(n)$, so that 
$$
\det(1) = a_1, \qquad \det(2) = a_1a_2 - a_0a_3,
$$
and so forth.   The requirement that the left side of
equation (\ref{poly}) be a Hurwitz polynomial
may then be succinctly stated as 
\beq\label{hurwitz}
\det(n) > 0, \qquad n=1,2 ... 5.
\eeq
This is the Routh-Hurwitz criterion.  The criterion (\ref{stbcond}) in
the text amounts to the condition $a_5>0$, which follows directly from
the two demands $\det(4)>0$ and $\det(5)=a_5\times \det(4) > 0$.  We
shall now prove that the general Routh-Hurwitz criterion is satisfied
for our dispersion relation.

The first non-trivial requirement is
\beq
\det(2) = a_1a_2 - a_0a_3 > 0.
\eeq
The determinant is easily calculated,
\beq
\det(2) = \kkz \frac{2\chi T}{5P} (\bb{k\cdot b})^2 \epsilon,
\eeq
and is, indeed, positive.

The calculation for $\det(3)$ is a bit more complicated,
\beq
\det(3) = a_3\, \det(2) - a_1 
\left| \begin{array} {cc}
a_1 & a_0 \\
a_5 & a_4
\end{array} \right|,
\eeq
which simplifies to 
\beq
\det(3) =  \kkz \left( \frac{2\chi T}{5P}\right)^2 \left( \bb{k\cdot b}\right)^4
\epsilon \left[ a_2 - \epsilon - (\bb{k\cdot v_A})^2 \right]  
\eeq
But the term in square brackets is just
\beq
(a_2 - \delta -\epsilon) + 4\Omega^2 + (\bb{k\cdot v_A})^2\left( \frac{k^2}{k_Z^2} -1
\right).
\eeq
According to equation (\ref{aaa}),
the sum of the first three grouped terms must be positive if $a_5 > 0$,
and the remaining two terms are each manifestly positive.  Hence $\det(3)
> 0$.

The calculation for $\det(4)$ proceeds as follows:
\beq\label{det4}
\det(4) = a_4\, \det(3) - a_5
\left| \begin{array} {ccc}
a_1 & a_0  & 0\\
a_3 & a_2  & a_0\\
a_5 & a_4  & a_2
\end{array} \right|,
\eeq
The determinant cofactor of $a_5$ in the above may be expanded and simplified,
reducing to
\beq
\left| \begin{array} {ccc}
a_1 & a_0  & 0\\
a_3 & a_2  & a_0\\
a_5 & a_4  & a_2
\end{array} \right| = 
\kkz \left( \frac{2\chi T}{5P}\right) \left( \bb{k\cdot b}\right)^2\epsilon \left[
a_2 - (\bb{k\cdot v_A})^2 \right]
\eeq
Using this in equation (\ref{det4}) for $\det(4)$, after some algebraic excursion
we find
\beq
\det(4) = \kkz \left( \frac{2\chi T}{5P}\right)^2 \left( \bb{k\cdot b}\right)^4
(\bb{k\cdot v_A})^2 \epsilon^2 \left[
(\bb{k\cdot v_A})^2 \left( \frac{k^2}{k_Z^2} - 1\right) + 4 \Omega^2 
\right],
\eeq
which is manifestly positive.  Finally, as noted, $\det(5) = a_5 \,
\det(4) > 0$, if $a_5 > 0$.  We have therefore shown that the condition
$a_5 >0$, equation (\ref{stbcond}) in the text, ensures that the real
parts of the roots of the polynomial in equation (29) all lie in the
left half of the complex $\sigma$ plane, and that the criteria
(\ref{hoil1}) and (\ref{hoil2}) are necessary and sufficient for the
convective-rotational stability of a hot plasma.  QED.

\newpage
\begin{figure}
\epsscale{0.40}
\plotone{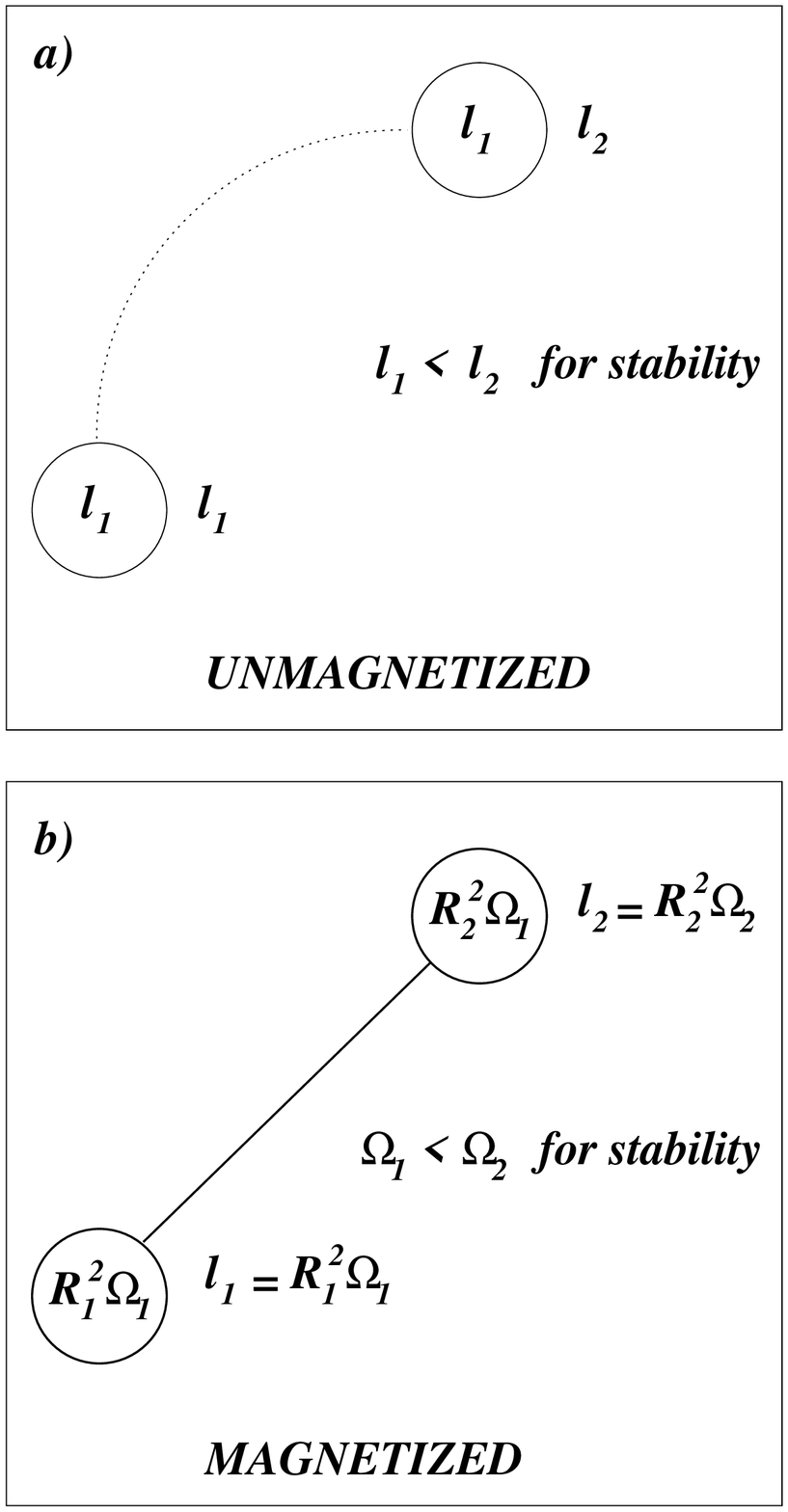}
\caption{
(a).~In an unmagnetized disk, specific angular momentum is
conserved.  Stability may be determined by comparing the angular
momentum of an outwardly displaced fluid element ($l_1$) with that of
its surroundings ($l_2$).
If $l_1<l_2$, the displaced fluid element has
less angular momentum than it needs to remain in its new orbit, and it
drops back.  This is equivalent to the Rayleigh condition $dl/dR> 0$.
(b).~In a magnetized disk, it is the angular velocity of
tethered fluid elements that tends to be conserved near
marginal stability.  Hence, if $\Omega_1<\Omega_2$, the
displaced fluid element has less angular momentum than it
needs to remain in its new orbit.  This corresponds to the
MRI stability condition, $d\Omega/dR>0$.}

\end{figure}

\begin{figure}
\epsscale{0.40}
\plotone{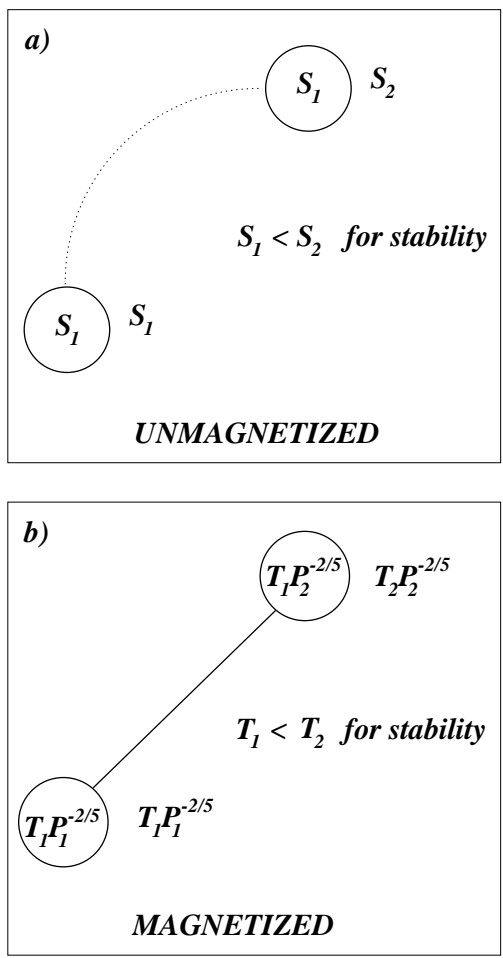}
\caption{ A process similar to that depicted in fig.~[1]
occurs in a thermally stratified disk.
(a).~ In an unmagnetized layer, entropy is conserved.
Stability may be determined by comparing the entropy
of an outwardly displaced fluid element ($S_1$) with that of
its surroundings ($S_2$).
If $S_1<S_2$, the displaced fluid element has
less entropy than it needs to remain in its new orbit, and it
drops back.  This is equivalent to the Schwarzschild condition
of increasing upward entropy.
(b).~In a magnetized conducting layer, it is the temperature of
tethered fluid elements that tends to be unchanged near marginal
stability.  As in (a), if $S_1<S_2$, the displaced fluid element has
less entropy than it needs to remain in its new orbit.  But since
$S\propto\ln TP^{-2/5}$, and pressure balance is maintained, this
corresponds to a thermoclinic stability condition: the background
temperature must increase upwards.}

\end{figure}

\begin{figure} \epsscale{1.0} \plotone{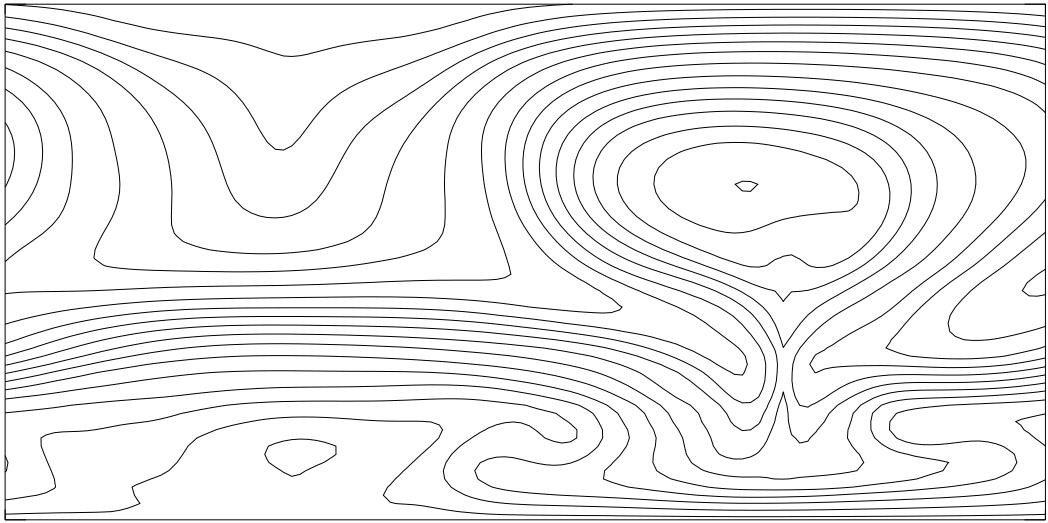}
\caption{Development
of thermoclinic instability in a Schwarzschild-stable layer.
Magnetic lines of force are shown after one Alfv\'en crossing time,
initial seeding with rms 1\% random initial vertical velocity
perturbations.  Initial thermal energy density is 1600 times magnetic;
initial field lines are isothermal and horizontal; vertical grid runs from
$z=1$ to $2$, initial temperature profile is $1/z$, gravitational field
is $1/z^2$; $\chi$ is 0.05; grid is $128 \times 64$.
The interpenetration of cool and warm blobs appears very
similar to classical convective instability.} \end{figure}

\end{document}